\documentstyle[preprint,aps,epsf]{revtex}                  
\global\let\epsfloaded=Y 
\begin{document}
\pagestyle{empty}                                      
\preprint{
\font\fortssbx=cmssbx10 scaled \magstep2
\hbox to \hsize{
\hfill $
\vtop{
 \hbox{ }}$
}
}
\draft
\vfill
\title{A new parametrization of the neutrino mixing matrix\\
for 
neutrino oscillations}
\vfill
\author{Virendra Gupta$^1$ and Xiao-Gang He$^2$}
\address{\rm $^1$ Departmento de Fisica Aplicada, CINVESTAV-Unidad Merida\\
A.P. 73 Cordemex 97310 Merida, Yucatan, Mexico\\
\rm $^2$ Department of Physics, National Taiwan University,
Taipei, Taiwan 10764, R.O.C.}

%
%
\vfill
\maketitle
\begin{abstract}
In this paper we study three active neutrino oscillations, favored
by recent data from SuperK and SNO, using a new parametrization of
the lepton mixing matrix $V$ constructed from
a linear combination of the unit matrix $I$, 
and a hermitian unitary matrix $U$, that is, 
$V = \cos\theta I + i\sin \theta U$. There are only three
real parameters in $V$ including the parameter $\theta$. 
It is interesting to find that experimental 
data on atmospheric neutrino dictates the angle $\theta$ to be $\pi/4$
such that the $\nu_\mu$ and $\nu_\tau$ mixing is maximal. 
The solar neutrino problem is solved via the MSW effect with a small 
mixing angle, with $U$ depending on one small parameter $\epsilon$.
The resulting mixing matrix with just two parameters ($\theta$ and $\epsilon$) 
predicts that the oscillating probabilities for
$\nu_e\to \nu_\mu$ and $\nu_e \to \nu_\tau$ to be equal and of the order
$2\epsilon^2 = (0.25\sim 2.5)\times 10^{-3}$. The measurement of CP
asymmetries at the proposed Neutrino Factories would also provide
a test of our parametrization.
\end{abstract}
%
%
%
%
\pagestyle{plain}

Flavor mixing in both quark and lepton sectors is one of the mysteries of
particle physics. There are abundant experimental data which show that 
different quark generations mix and so do lepton generations. However,
theoretical understanding of the mixing is very poor. The minimal 
Standard Model (SM) can accommodate quark mixing through the 
Cabibbo-Kobayashi-Maskawa mixing matrix\cite{1}. 
For mixing in the lepton sector,
one has to extend the minimal SM, either by introducing an new scalar particle 
to provide Majorana masses for neutrinos and therefore mixing or by 
introducing right-handed neutrinos to provide neutrinos with Dirac mass, or 
a combination of these two. The mixing phenomena in the lepton sector are
most prominent in neutrino oscillations and can be described by the 
Maki-Makagawa-Sakata mixing 
matrix\cite{2}. These proposals can accommodate mixing phenomena
and can be made consistent with experimental data, but there is no 
understanding
why the matrix elements in relavent mixing matrix have the determined 
values.  

In this paper
assuming three generations of neutrinos, 
we consider another way of looking at the mixing matrix problem 
by dividing the mixing 
matrix into two pieces. One of them is proportional to the unit matrix $I$
which dose not cause any mixing,
and the other is proportional to 
a hermitian unitary matrix $U$ which is responsible for flavor mixing. 
The mixing matrix $V$ 
is thus parametrized 
as $V=\cos \theta I + i \sin\theta U$. This way of looking at the problem 
has the advantage that the parameter $\theta$ determines the relative weight 
of the diagonal elements and off diagonal elements in a non-trivial manner.
Mathematically, $U$ depends on four real parameters, but due to the rephasing
invariance of $V$, there are actually just two real independent parameters
in $U$. Similar proposal has been made
for the quark mixing matrix which is compatible with experimental data within
allowed errors\cite{4}. 
Here we will concentrate on the lepton sector to
analyze mixing phenomena in neutrino oscillations. 
Neutrino oscillations can provide natural explanations to the
atmospheric and solar neutrino data. Recent data from SuperK\cite{5} and 
SNO\cite{6}
favor mixing of active neutrinos. We therefore assume three active neutrino
mixing in our analysis. We note that the 
LSND\cite{7} data cannot
be accommodated in this analysis. 
The required parameter space to exlain the atmospheric and solar neutrino data
can provide stringent constraints on the parameters
in the mixing matrix. 
Using atmospheric neutrino data, 
we find that the linear combination angle $\theta$ 
is determined to be $\pi/4$ indicating that the diagonal and off diagonal 
parts are equally important. The solar neutrino problem can be solved by the
small mixing angle (SMA) solution with MSW effect. 
This ansatz also gives interesting predictions for long base line 
neutrino experiments and 
CP violation in neutrino oscillations.  

We now provide a detailed analysis of the parametrization proposed here.
Unitarity of $V$ requires

\begin{eqnarray}
&&VV^\dagger = I = \cos^2\theta I+ \sin^2\theta UU^\dagger 
+i\sin\theta \cos\theta (U-U^\dagger),\nonumber\\
&&V^\dagger V = I = \cos^2\theta I + \sin^2\theta U^\dagger U + i\sin\theta 
\cos\theta(U-U^\dagger).
\end{eqnarray}
In general $V$ and $U$ are non-trivially related. 
However, if $U$ is independent of the linear combination angle $\theta$, 
then $U$ has to be unitary and hermitian.
The $V$ so obtained is more restrictive than a general one.

The matrix $U$ can be parametrized as

\begin{eqnarray}
U = \left (
\begin{array}{ccc}
-1+2|c|^2&-2b^* c&-2a^* c^*\\
-2bc^*&-1+2|b|^2&-2ab^*\\
-2ac&-2a^*b&-1+2|a|^2
\end{array}
\right ),
\end{eqnarray}
where the magnitudes and the phases of $a$, $b$ and $c$ satisfy 
the constraints $|a|^2+|b|^2+|c|^2 = 1$
and $\phi_a-\phi_b+\phi_c = \pi/ 2$, respectively. $U$ depends, in general,
on two independent moduli and two independent phases at it stands. However the
mixing matrix $V$ is rephasing invariant, so the two
phases in $U$ can be absorbed into lepton phases. 
$V$ depends on only 3 real parameters, $\theta$ and the two moduli in $U$.

To determine the parameters in the mixing matrix $V$, we first
consider data for atmospheric neutrino from 
SuperK
experiment\cite{5}. To fit the data 
the muon neutrino $\nu_\mu$ has to
oscillate into another type of neutrino $\nu_x$ 
with mixing angle $\theta_{2x}$ such that 
$sin^22\theta_{2x} >0.88$ which leads to 
$\theta_{2x}\approx \pi/4$ and also the mass squared difference
$|\Delta m^2_{2 x}| = |m^2_2 -m^2_x|$ in the range  
$(1.5\sim 5)\times 10^{-3}$ eV$^2$. Here
the numbers 1, 2 and 3 correspond to $\nu_e$, $\nu_\mu$ 
and $\nu_\tau$, respectively. 
For three generations of neutrinos,
$\nu_\mu$ can only oscillate into a tauon neutrino $\nu_\tau$ to fit the data.  
Assuming that electron neutrino $\nu_e$ 
does not have significant mixing with muon and
tauon neutrinos, then
$|V_{11}|$ is fixed to be $1$, while $|V_{22}|$, $|V_{23}|$, $|V_{32}|$ 
and $|V_{33}|$ are 
fixed to be $1/\sqrt{2}$. This leads to the following unique solution for
$\theta$ and $U$,

\begin{eqnarray}
&&\theta = {\pi \over 4},\;\;\;\;
a={1\over \sqrt{2}},\;\;\;\;b={1\over \sqrt{2}},\;\;\;\;c = 0,\nonumber\\
&&U = \left (
\begin{array}{lll}
-1&0&0\\
0&0&-1\\
0&-1&0
\end{array}
\right ).
\end{eqnarray}
The mass squared difference $|\Delta m^2_{23}|$ is constrained to be in 
the range $( 1.5 \sim 5)\times 10^{-3} \mbox{eV}^2$ at 90\% C.L..
The mixing angle $\theta = \pi/4$ is a reflection of the fact that the 
mixing of the $\nu_\mu$ and $\nu_\tau$ is maximal. 
It is interesting that this value of $\theta$ also implies that the purely
diagonal and the non-diagonal pieces have equal weight, that is,
$V = (I + iU)/\sqrt{2}$.

The mixing matrix obtained above can not produce oscillations of electron 
neutrino into other neutrinos and
therefore can not provide a solution to the solar neutrino problem. 
If the solar neutrino problem is due to neutrino oscillation like the
solution to the atmospheric neutrino problem, then the 
electron neutrino has to mix 
with other neutrinos such that it can oscillate into them to cause
the deficit seen in the measurements on earth. 
One has to modify the above mixing matrix such that 
solutions for the solar neutrino problem can be obtained. 

To this end
we make a small perturbation to the mixing matrix by 
introducing a non-zero but small value for 
$c$ such that $V_{12,23}$ become non-zero to allow
$\nu_e$ to oscillates into $\nu_\mu$ and/or $\nu_\tau$. In fact, in our 
parametrization, a non-zero value for $c$ implies that both $V_{12}$ and
$V_{13}$ are non-zero. Whether the solar neutrino problem is solved by
oscillation of 
$\nu_e$ into $\nu_\mu$ or $\nu_\tau$ depends on the mass squared 
differences of the  neutrino masses which will be discussed later. 
Denote the perturbations on $a$, $b$ and $c$ as 
$a = (1+\epsilon_a)/\sqrt{2}$,   
$b = (1+\epsilon_b)/\sqrt{2}$, and $c = i\epsilon$. If $\epsilon$
is a real number, $\epsilon_{a,b}$ are also required to be real from the
constraint on the phases. 
A simple choice, satisfying all constraints, is to take $c=i\epsilon$
with $a = \sqrt{1-2\epsilon^2}/\sqrt{2}$ and $b=1\sqrt{2}$. 
this gives

\begin{eqnarray}
V = {1\over \sqrt{2}}\left ( \begin{array}{ccc}
1-i(1-2\epsilon^2)&\sqrt{2}\epsilon&
-\sqrt{2}\epsilon \sqrt{1- 2\epsilon^2}\\
-\sqrt{2}\epsilon&1&
-i\sqrt{1-2\epsilon^2}\\
\sqrt{2}\epsilon\sqrt{1- 2\epsilon^2}&-i\sqrt{1-2\epsilon^2}
&1-i2\epsilon^2
\end{array} 
\right ).
\end{eqnarray}
There are other choices for $\epsilon_{a,b}$ which can also satisfy 
$|a|^2+|b|^2+|c|^2 = 1$, such as the general case
$a = \sqrt{1- 2(1+\alpha) \epsilon^2}/\sqrt{2}$ and
$b = \sqrt{1+2\alpha \epsilon^2}/\sqrt{2}$, with $\alpha$ a number
of order one so that $\epsilon_{a,b}$ are small. 
However, to the leading order in $\epsilon^2$ the general case gives 
the same
results for neutrino oscillations. Without loss of generality, 
we will carry out the rest of the analysis using $V$
given above.

There are four types of solution to the solar neutrino problem\cite{8}. 
Three of them
are based on the MSW mechanism\cite{9}. They are
the large mixing angle (LMA), 
the long wavelength (LOW) large mixing
angle, and the small mixing angle (SMA) solutions. There is one solution which
does not depend on the MSW effect, the ``just so" vacuum (VAC) oscillation
solution. Recent data from SuperK favors the LMA solution\cite{8}. 
However, at present it
is premature to rule out the other three solutions. Among the four 
solutions, our parametrization can only accommodate the SMA solution. 
Fitting data we determine the parameter $\epsilon^2$ and the mass squared 
difference of $\nu_e$ and the neutrino oscillated into $\nu_x$ to be

\begin{eqnarray}
&&\epsilon^2 = (0.125\sim 1.25 )\times 10^{-3},\nonumber\\
&&|\Delta m^2_{1x}| = (0.35 \sim 1)\times 10^{-5}\mbox{eV}^2.
\end{eqnarray}
Because the mixing matrix elements $V_{12}$ and $V_{13}$ 
to the leading order in $\epsilon$ are equal, present constraints on various
neutrino oscillation data can allow
the neutrino $\nu_x$ to be $\nu_\tau$ or $\nu_\mu$ depending on the
pattern of the neutrino masses. There are two scenarios: 
a) if $\nu_e$ oscillates into $\nu_\tau$, the neutrino mass hierarchy will be
$m_{\nu_\mu} > m_{\nu_\tau} > m_{\nu_e}$; and b) if
$\nu_e$ oscillates into $\nu_\mu$, then $m_{\nu_\tau} > m_{\nu_\mu} > 
m_{\nu_e}$. 

The mass hierarchy patterns for the cases a) and b) have different 
implications in general 
for long base line experiments. The probability for 
a $l$ type of neutrino to oscillate into a $k$ type of neutrino in 
vacuum is given by

\begin{eqnarray}
P(l\to k) &=& \delta_{lk} - 4 \sum_{i<j} Re(V^*_{ki}V_{li}V^*_{lj}V_{kj})
\sin^2({1.27\Delta m^2_{ij}\over E} L)\nonumber\\
&-&2\sum_{i<j}
Im(V^*_{ki}V_{li}V_{lj}V^*_{kj})\sin({2.54\Delta m^2_{ij}\over E} L),
\end{eqnarray}
where $\Delta m^2_{ij}$ is in
eV$^2$, $E$ in GeV and $L$ in $km$. 
The first two terms are CP conserving and symmetric in $l$ and $k$ while the
last term is CP violating and anti-symmetric in $l$ and $k$. For anti-neutrino
oscillation the last term will change sign.

In our new parametrization, to order $\epsilon^2$ 
we find the following oscillation probabilities

\begin{eqnarray}
&&P(\nu_e\to \nu_e) = 1 - 4\epsilon^2 [\sin^2({1.27 \Delta m^2_{12}\over E} L)
+ \sin^2 (1.27{\Delta m^2_{13}\over E}L)],\nonumber\\
&&P(\nu_\mu\to \nu_\mu)= P(\nu_\tau \to \nu_\tau) 
\nonumber\\
&&= 1 - 2\epsilon^2 [\sin^2({1.27 \Delta m^2_{12}\over E} L)
+ \sin^2 ({1.27\Delta m^2_{13}\over E}L)] 
-(1-2\epsilon^2) \sin^2({1.27\Delta m^2_{23}\over E}L),\nonumber\\
&&P(\nu_e\to \nu_\mu)
=P(\nu_\tau \to \nu_e) 
=2\epsilon^2 [\sin^2({1.27 \Delta m^2_{12}\over E} L)
+ \sin^2 ({1.27\Delta m^2_{13}\over E}L)] + P^-,\nonumber\\
&&P(\nu_\mu\to \nu_\tau)
=(1-2\epsilon^2) \sin^2({1.27\Delta m^2_{23}\over E}L)
+ P^-,\nonumber\\
&&P^- =- 4\epsilon^2 \sin({1.27\Delta m^2_{12}\over E}L) 
\sin({1.27\Delta m^2_{13}\over E}L) \sin({1.27\Delta m^2_{23}\over E}L).
\end{eqnarray}

For the large $\nu_\mu \to \nu_\tau$ oscillation, further tests can be
carried out to check the oscillation mechanism at the long base line 
neutrino experiments\cite{10} at K2K, Minos and Opera in a controlled fashion.
In these experiments, it may also be possible to 
carry out some tests for the predictions
of the ansatz proposed here.
In both cases a) and b), it is feasible to have the right combination of 
$L/E$ such that $1.27\Delta m^2_{12}L/E$ or $1.27\Delta m^2_{13}L/E$ 
are of order unity so that oscillations between $\nu_e$ and $\nu_\mu$, and
$\nu_e$ and $\nu_\tau$ can be observed. 
The long base line neutrino oscillation experiments 
K2K, Minos and Opera have $L$ around 250km 730km, and 730km with
$E$ around $\sim 1.4$ GeV, $\sim 3,\; 7,\; 15$ GeV, and $\sim 17$ GeV, 
respectively.
It is possible using $E_{\nu_\mu} = 3$ GeV from Minos to have 
$1.27\Delta m^2_{12}L/E$ or $1.27\Delta m^2_{13}L/E$ 
as large as $\pi/2$.
The maximal probability at the far end of the long base line detector 
for the appearing
of $\nu_e$ can be as large as $2\epsilon^2 = 
(0.25\sim 2.5)\times 10^{-3}$ from $\nu_\mu \to \nu_e$ oscillation. 
The appearing probability of 
$\nu_\mu$
can be in the range $(0.25\sim 2.5)\times 10^{-3}$ 
from $\nu_e \to \nu_\mu$ oscillation. Future experimental data
will provide important information about the mixing matrix. Unfortunately,
the cases a) and b), to the leading order, give identical predictions 
for the above mentioned experiments and can not be distinguished.

Another interesting prediction of the new mixing matrix parametrization is
CP violation in neutrino oscillations. Tests of 
CP violation in neutrino oscillations
may be feasible at Neutrino Factories\cite{11,12}.
Proposed Neutrino Factories have a high intensity muon storage ring with
long sections along which the muons decay, $\mu^- \to \nu_\mu e^- \bar \nu_e$,
to deliver high intensity $\nu_\mu$ and $\bar \nu_e$ neutrino
beams. With these neutrino beams one can carry out disappearing and appearing
neutrino experiments to test different neutrino mixing scenarios.
Defining CP violating asymmetry,

\begin{eqnarray}
A(\nu_l\to \nu_k) = {P(\nu_l \to \nu_k) - P(\bar \nu_l \to \bar \nu_k)
\over P(\nu_l \to \nu_k) + P(\bar \nu_l \to \bar \nu_k)},
\end{eqnarray}
we have

\begin{eqnarray}
&&A(\nu_e \to \nu_\mu) = A(\nu_\tau\to \nu_e) = -A(\nu_\mu\to \nu_e)
= -A(\nu_e \to \nu_\tau)\nonumber\\
&&\nonumber\\
&&=-2{\sin({1.27\Delta m^2_{12} \over E}L)\sin({1.27\Delta m^2_{13}\over E}L)
\sin({1.27\Delta m^2_{23}\over E}L)
\over \sin^2({1.27\Delta m^2_{12}\over E}L) + \sin^2({1.27\Delta m^2_{13}
\over E}L)}.
\end{eqnarray}
In the above we have given the modes which may have large CP violation.
In principle $A(\nu_i\to \nu_i)$, $A(\nu_\mu\to \nu_\tau)$ and
$A(\nu_\tau \to \nu_\mu)$ can have non-zero values. However there is an 
additional suppression factor of order $\epsilon^2$ for CP violation 
in these modes and they would be difficult to measure.

It is interesting to note that in the above expression the mixing parameter 
$\epsilon$
has canceled out, and also that all 
the above modes have CP violation of the same magnitude. 
Therefore uncertainties associated with the mixing 
parameters do not enter and allow clear interpretations. 
These unique features can provide good tests for 
the new parametrization studied here. 

Let us now study what value $A(\nu_\mu\to \nu_e)$ can have.
One can tune $L/E$ to maximize the detection probability. 
Tuning $1.27\Delta m^2_{21}L/E = \pi/2$
for case a) and substituting $1.27\Delta m^2_{31}L/E = (\Delta m^2_{31}
/\Delta m^2_{21}) \pi/2$, we have

\begin{eqnarray}
A(\nu_e \to \nu_\mu) &=&- 2 \sin({\Delta m^2_{31}\over \Delta m^2_{21}}
{\pi\over 2}) \approx - 2{\Delta m^2_{31}\over \Delta m^2_{21}}{\pi\over 2}
\nonumber\\
&=&- (0.6\sim 0.6)\times 10^{-2}.
\end{eqnarray}
In case b), one would get the same magnitude, but opposite sign. This
fact can be used to distinguish the cases a) and b).
To have a $3\sigma$ detection of CP asymmetry, one would need 
$(10^{10} -10^{8})$ neutrino oscillation events. 

When neutrino beams travel through
the earth to reach the other side to be detected, due to the 
fact that the earth is composed of
matter and not anti-matter, the MSW effect for the neutrino and 
anti-neutrino beams is different. 
This results in additional asymmetry. To have some 
idea about the additional asymmetry due to matter effects, we consider 
case b) using 
$m_1 = 0$, $m_2 = 2.5\times 10^{-3}$ eV, $m_3 = 6\times 10^{-2}$ eV and 
$\epsilon^2 = 5\times 10^{-4}$ as the input vacuum values for the neutrino 
masses and mixing for illustration. The matter effect is proportional to the
electron density $N_e$ in the earth. The parameter that 
enters in the calculation is
$A = 2\sqrt{2} G_F N_e E_\nu$. 
For neutrino oscillation, $A$ takes the positive sign and for anti-neutrino
oscillation it takes the minus sign. 
For the typical neutrino energy of the Neutrino Factories and the electron 
density in the earth, $A$ is in the range of $10^{-7} \sim 10^{-2}$ 
eV$^2$\cite{13}. 
We find that in vacuum for the above values of parameters, 
the CP asymmetry is predicted to be 0.0055. But with a non zero $A$, the
result can be very different. For example for $A=10^{-3}$ eV$^2$, the asymmetry 
would be 0.035. 
One should note that the asymmetry caused by the matter effect does not imply
CP violation in the mixing matrix. To pin down the effect due to CP violation, 
one needs to have a good understanding of the profile of the
earth to eliminate the asymmetry due to the MSW effect on neutrino oscillations.
This is very difficult. However, we would like to point out that
$A(\nu_l\to \nu_k) = [P(\nu_l\to \nu_k) - P(\bar \nu_l \to \bar \nu_k)]
/[P(\nu_l\to \nu_k) + P(\bar \nu_l\to \bar \nu_k)]$ can still provide 
important information about the neutrino mass hierarchy. Because in case a) and
case b) the signs for $A(\nu_e\to \nu_\mu)$ and 
$A(\nu_e\to \nu_\tau)$ are different, and
matter effects do not change the signs of the asymmetries 
for $A$ in the range $10^{-7}
\sim 10^{-2}$ eV$^2$.

We have considered a new parametrization for neutrino mixing matrix $V$ using 
the ansatz $V= \cos\theta I + i \sin\theta U$ with the value $\theta = \pi/4$
which gives equal weight to the diagonal ($I$) and non-diagonal ($U$) pieces 
in $V$. This value of $\theta$ is a natural consequence of the 
atmospheric neutrino data which requires maximal $\nu_\mu \to \nu_\tau$ 
oscillation. The matrix
$U$ depends on only one small parameter $\epsilon\sim 10^{-2}$ which gives the
SMA solution to the solar neutrino problem. It also predicts that
$P(\nu_e\to \nu_\mu) = P(\nu_\tau \to \nu_e)$ to the leading order.

We find that to order $\epsilon^2$, the CP asymmetry $A(\nu_e \to \nu_l)$ 
is independent of $\epsilon$. Moreover, we find that the sign of
$A(\nu_e\to \nu_\mu)$ and $A(\nu_e\to \nu_\tau)$ could be used to distinguish
between the two possible mass hierarchies,
$m_{\nu_\mu} > m_{\nu_\tau} > m_{\nu_e}$ and 
$m_{\nu_\tau} > m_{\nu_\mu}> m_{\nu_e}$ permitted by the SMA solution to the
solar neutrino problem and other oscillation data, even though 
matter effects may 
affect the magnitude of the asymmetry significantly. 
These asymmetries could be detected at Neutrino Factories. 

It is remarkable that our mixing matrix $V$ which contains only two parameters
($\theta$ and $\epsilon$)  can explain the
atmospheric and solar neutrino data. CP asymmetry measurements in 
neutrino oscillations could provide future tests 
of our parametrization.

This was supported in part
by National Science Council under grant NSC 89-2112-M-002-058 and in part
by the Ministry of Education Academic Excellence Project 89-N-FA01-1-4-3.
VG would like to thank the Department of Physics, National Taiwan University
for hospitality while this work was completed.

\end{document}